\documentstyle[12pt]{article}

\font\tenof=msym10
\font\bdi=cmmib10 at 12 pt

\def\bi#1{\hbox{\bdi #1\/}}

\renewcommand{\theequation}{\arabic{section}.\arabic{equation}}
\newcommand{\rb}{\bi{r}}
\newcommand{\pb}{\bi{p}}
\newcommand{\xb}{\bi{x}}
\newcommand{\Lb}{\bi{L}}
\newcommand{\Sb}{\bi{S}}
\newcommand{\Jb}{\bi{J}}
\newcommand{\alphab}{\bi{\char'13}}

\newcommand{\etab}{\bi{\char'21}}
\newcommand{\xib}{\bi{\char'30}}
\newcommand{\varphib}{\bi{\char'47}}
\newcommand{\sigmab}{\bi{\char'33}}
\newcommand{\Deltab}{\mathbf{\Delta}}
\newcommand{\C}{{\cal C}}
\newcommand{\m}{\phantom{-}}

\def\case#1#2{{\textstyle{#1\over #2}}}
\def\I{\mbox{\tenof I}}
\def\R{\mbox{\tenof R}}

\begin{document}
\baselineskip=22pt plus 1pt minus 1pt
%
%
{\parindent=0cm
{\Large \bf Supersymmetry and superalgebra for the two-body\\
system with a Dirac oscillator interaction}

\vspace{2cm}
{\large \bf M Moshinsky$^*$\footnote[2]{Member of El Colegio Nacional}, C
Quesne$^{\ddag}$\footnote[4]{Directeur de recherches FNRS}\footnote[5]%
{e-mail: cquesne@ulb.ac.be} and Yu F Smirnov$^*$}

\medskip
$^*$ Instituto de F\'\i sica, UNAM, Apdo.\ Postal 20-364, 01000 M\'exico DF,
Mexico

$^{\ddag}$ Physique Nucl\'eaire Th\'eorique et Physique Math\'ematique,
Universit\'e Libre de Bruxelles, Campus de la Plaine CP229, Boulevard du
Triomphe, B-1050 Brussels, Belgium

\vspace{5cm}
Short title: Two-body system with Dirac oscillator interaction

\bigskip
PACS numbers: 02.20.+b, 03.65.Fd, 11.30.-j}
\newpage
%
%
\noindent
{\bf Abstract.} Some years ago, one of the authors~(MM) revived a concept to
which
he gave the name of single-particle Dirac oscillator, while another~(CQ) showed
that it corresponds to a realization of supersymmetric quantum mechanics. The
Dirac oscillator in its one- and many-body versions has had a great number of
applications. Recently, it included the analytic expression for the eigenstates
and
eigenvalues of a two-particle system with a new type of Dirac oscillator
interaction of frequency~$\omega$. By considering the latter together with its
partner corresponding to the replacement of~$\omega$ by~$-\omega$, we are
able to get a supersymmetric formulation of the problem and find the
superalgebra that explains its degeneracy.\par
\newpage
%
%
\section{Introduction}
A Dirac equation with an interaction linear in the coordinates was considered
long
ago~\cite{0} and revived more recently~\cite{1} with the name of Dirac
oscillator as, when reduced to the large component, it corresponds to a
standard
harmonic oscillator with a very strong spin-orbit term.\par
%
%
The concept gave rise to a large number of papers concerned with its different
aspects~\cite{2}. Two of the present authors (MM and CQ) were particularly
interested in the symmetry Lie algebra~\cite{3} that explains the degeneracy of
the spectrum of the one-body Dirac oscillator, and later one of them (CQ)
showed~\cite{4} that the problem corresponds to a realization of supersymmetric
quantum mechanics~\cite{4a}.\par
%
%
One of the authors (MM) and his collaborators were particularly concerned with
two- and three-body systems with Dirac oscillator interactions, because in the
former case it gave insight into the quark-antiquark system and the mass
formula
for mesons~\cite{5,6,7}, while in the latter it could be applied to the
three-quark
system and the mass formula of baryons~\cite{8,9}.\par
%
%
In all the extensions to more than one particle, the analysis was made in terms
of
a single Poincar\'e invariant equation of the type employed by Barut and his
collaborators~\cite{10}. The most usual approach to the $n$-body problem is
through the use of $n$ separate equations but that satisfy appropriate
compatibility conditions~\cite{11,12,13}.\par
%
%
In a recent paper~\cite{14}, Del Sol Mesa and Moshinsky showed that the two
approaches mentioned in the previous paragraph are equivalent, at least for a
two-body problem with a modified type of Dirac oscillator interaction of
frequency~$\omega$. This last problem, with a fully Poincar\'e invariant
formulation, has, in the centre-of-mass frame, very simple analytic
expressions for its eigenvalues and eigenstates, and it displays an
extraordinary
accidental degeneracy. In the present paper, we shall prove that when the
problem
of frequency~$\omega$ is taken together with that of its supersymmetric
partner,
corresponding to the replacement of~$\omega$ by~$-\omega$ ( and referred to in
the following as the Dirac oscillator with negative frequency), a superalgebra
explains the degeneracies that appear.\par
%
%
Before entering into the objective of the paper, indicated at the end of the
last
paragraph, we note that in the one-body case, Quesne~\cite{4} considered as
supersymmetric partners the large components of particle and antiparticle wave
functions. It is equivalent though~\cite{5} to consider as supersymmetric
partners
the large components of the wave functions of frequency~$\omega$
and~$-\omega$, and it will be this point of view that will be generalized in
the
present paper.\par
%
%
In section~2, we determine the spectrum and the eigenstates of the two-body
Dirac oscillator and of its partner with negative frequency. In section~3, we
then
reformulate the two-body problem in the framework of supersymmetric quantum
mechanics and obtain the superalgebra explaining the spectrum degeneracies.
Finally, section~4 contains the conclusion.\par
%
%
\section{The two-body problem with a new type of Dirac oscillator interaction}
\setcounter{equation}{0}
In ref.~\cite{14}, a two-body problem with a new type of Dirac oscillator
interaction was introduced through a Poincar\'e invariant equation. In the
centre-of-mass frame, the latter becomes
\begin{equation}
  \left[(\alphab_1 - \alphab_2) \cdot \left(\pb \mp i \case{\omega}{2} \rb
\beta_1
  \beta_2 \gamma_{51} \gamma_{52}\right) + \beta_1 + \beta_2\right] \psi^{\pm}
  = E \psi^{\pm}                \label{eq:3.8}
\end{equation}
where $\rb$, $\pb$ are defined in terms of the coordinates $\xb_1$,~$\xb_2$,
and
momenta $\pb_1$,~$\pb_2$ of the two particles as
\begin{equation}
  \rb \equiv \xb_1 - \xb_2 \qquad \pb \equiv \case{1}{2} (\pb_1 - \pb_2)
  \label{eq:3.9}
\end{equation}
while $\alphab_1$ and $\alphab_2$ are the direct product matrices
\begin{equation}
  \alphab_1 = \left(\begin{array}{cc}
                                  0                & \sigmab_1 \\
                                  \sigmab_1 & 0
                               \end{array}\right) \otimes
                     \left(\begin{array}{cc}
                                  I  & 0 \\
                                  0 & I
                               \end{array}\right) \qquad
  \alphab_2 = \left(\begin{array}{cc}
                                  I  & 0 \\
                                  0 & I
                               \end{array}\right) \otimes
                      \left(\begin{array}{cc}
                                  0                & \sigmab_2 \\
                                  \sigmab_2 & 0
                               \end{array}\right)
  \label{eq:3.4}
\end{equation}
and similarly for $\beta_1$, $\beta_2$, $\gamma_{51}$, and~$\gamma_{52}$. Note
that as we want to simultaneously analyse the Dirac oscillators
for frequencies~$\omega$ and~$-\omega$, in eq.~(\ref{eq:3.8}) we have written a
$\mp$~sign before the term proportional to~$\rb$, and we have appended a
corresponding $\pm$~superscript to~$\psi$.\par
%
%
It is now a question of discussing the eigenstates~$\psi^{\pm}$
of~(\ref{eq:3.8})
and the corresponding eigenvalues of the energy. We first note that
from~(\ref{eq:3.4}), $\alphab_1$, $\alphab_2$, $\beta_1 \beta_2 \gamma_{51}
\gamma_{52}$, $\beta_1 + \beta_2$ can be written as $4\times 4$~matrices in the
form
\begin{eqnarray}
  \alphab_1 & = & \left(\begin{array}{cccc}
                                        0                & \sigmab_1 & 0
        & 0                \\
                                        \sigmab_1 & 0                & 0
        & 0                \\
                                        0                & 0                & 0
               & \sigmab_1 \\
                                        0                & 0                &
\sigmab_1 & 0
                                     \end{array}\right) \qquad
  \alphab_2 = \left(\begin{array}{cccc}
                                  0                & 0                &
\sigmab_2 & 0                \\
                                  0                & 0                & 0
         & \sigmab_2 \\
                                  \sigmab_2 & 0                & 0
  & 0                \\
                                  0                & \sigmab_2 & 0
  & 0
                              \end{array}\right) \nonumber \\
  \beta_1 \beta_2 \gamma_{51} \gamma_{52}
        & = & \left(\begin{array}{rrrr}
                              0 & 0  & 0  & I \\
                              0 & 0  & -I & 0 \\
                              0 & -I & 0  & 0 \\
                              I  & 0  & 0  & 0
                          \end{array}\right) \qquad
  \beta_1 + \beta_2 = \left(\begin{array}{rrrr}
                                               2 & 0 & 0 & 0  \\
                                               0 & 0 & 0 & 0  \\
                                               0 & 0 & 0 & 0  \\
                                               0 & 0 & 0 & -2
                                           \end{array}\right)
  \label{eq:3.10}
\end{eqnarray}
acting on a four-component vector
\begin{equation}
  \psi = \left(\begin{array}{c}
                         \psi_{11} \\
                         \psi_{21} \\
                         \psi_{12} \\
                         \psi_{22}
                     \end{array}\right)
  \label{eq:3.11}
\end{equation}
where for the moment we suppress the $\pm$~sign as we first deal with the case
of positive frequency~$\omega$.\par
%
%
If we now introduce the notation~\cite{14}
\begin{equation}
  a_{\pm} = (\sigmab_1 \cdot \pb) \pm i (\omega/2) (\sigmab_2 \cdot \rb)
  \qquad b_{\pm} = - (\sigmab_2 \cdot \pb) \pm i (\omega/2) (\sigmab_1 \cdot
  \rb)         \label{eq:3.12}
\end{equation}
equation~(\ref{eq:3.8}) can be written as
\begin{equation}
  \left(\begin{array}{cccc}
              (2-E) & a_- & b_+ & 0        \\
              a_+   & -E   & 0     & b_-     \\
              b_-   & 0     & -E   & a_+     \\
              0      & b_+  & a_- & (-2-E)
           \end{array}\right)
  \left(\begin{array}{c}
              \psi_{11} \\
              \psi_{21} \\
              \psi_{12} \\
              \psi_{22}
           \end{array}\right) = 0.          \label{eq:3.13}
\end{equation}
\par
%
%
The second and third rows of the matrix operator equation~(\ref{eq:3.13}) allow
us
to express $\psi_{21}$,~$\psi_{12}$ in terms of
$\psi_{11}$,~$\psi_{22}$, so that substituting them in the first and fourth
rows, we
obtain a $2\times 2$ matrix operator equation for the two components
$\psi_{11}$
and~$\psi_{22}$. Introducing then, as in previous work~\cite{15}, $\phi_1$
and~$\phi_2$ by the definitions
\begin{equation}
  \left(\begin{array}{c}
              \phi_1 \\
              \phi_2
           \end{array}\right)
  = \frac{1}{\sqrt{2}} \left(\begin{array}{rr}
                                              1 & -1 \\
                                              1 & 1
                                           \end{array}\right)
                                 \left(\begin{array}{c}
                                              \psi_{11} \\
                                              \psi_{22}
                                          \end{array}\right)
\label{eq:3.14}
\end{equation}
we finally obtain an equation of the form
\begin{equation}
  \left(\begin{array}{cc}
              A^+ - E^2 & 2E           \\
              2E           & B^+ - E^2
           \end{array}\right)
  \left(\begin{array}{c}
              \phi^+_1 \\
              \phi^+_2
           \end{array}\right) = 0        \label{eq:3.15}
\end{equation}
where
\begin{eqnarray}
  A^+ & \equiv & (a_- - b_+)(a_+ - b_-) = 4\omega \left[\Sb^2 + (\Sb \cdot
\etab)
         (\Sb \cdot \xib) + \Lb \cdot \Sb\right] \nonumber \\
        & = & 4\omega (\Sb \cdot \xib) (\Sb \cdot \etab) \label{eq:3.16} \\
  B^+ & \equiv & (a_- + b_+)(a_+ + b_-) = 4\omega \left[\hat N - (\Sb \cdot
\etab)
         (\Sb \cdot \xib) - \Lb \cdot \Sb\right] \nonumber \\
        & = & 4\omega (\Sb' \cdot \etab) (\Sb' \cdot \xib) \label{eq:3.17}
\end{eqnarray}
with creation~$\etab$ and annihilation~$\xib$ operators defined by
\begin{equation}
  \etab = \case{1}{\sqrt{2}} \left((\omega/2)^{1/2} \rb - i (\omega/2)^{-1/2}
  \pb\right) \qquad
  \xib = \case{1}{\sqrt{2}} \left((\omega/2)^{1/2} \rb + i (\omega/2)^{-1/2}
  \pb\right)      \label{eq:2.5}
\end{equation}
and
\begin{equation}
  \hat{N} = \etab \cdot \xib \qquad \Lb = \rb \times \pb = -i (\etab \times
\xib)
        \label{eq:2.9}
\end{equation}
while
\begin{equation}
  \Sb = \case{1}{2} (\sigmab_1 + \sigmab_2) \qquad \Sb' = \case{1}{2}
(\sigmab_1 -
  \sigmab_2).                \label{eq:3.18}
\end{equation}
\par
%
%
Note that we have added an upper sign~$+$ to $A$, $B$, $\phi_1$,~$\phi_2$ to
indicate that they belong to the equation where the frequency is~$\omega$.
{}From~(\ref{eq:3.8}), we see that the equation for negative
frequency~$-\omega$ can
be reduced to that for positive frequency~$\omega$ if we replace $\rb$
by~$-\rb$,
which from~(\ref{eq:2.5}) and~(\ref{eq:2.9}) implies that $\etab \to - \xib$,
$\xib
\to - \etab$, $\hat N \to \hat N + 3$, $\Lb \to - \Lb$.\par
%
%
Thus for the corresponding problem with frequency~$-\omega$, we have the
equation
\begin{equation}
  \left(\begin{array}{cc}
              A^- - E^2 & 2E           \\
              2E           & B^- - E^2
           \end{array}\right)
  \left(\begin{array}{c}
              \phi^-_1 \\
              \phi^-_2
           \end{array}\right) = 0        \label{eq:3.19}
\end{equation}
where
\begin{eqnarray}
  A^- & = & 4\omega \left[\Sb^2 + (\Sb \cdot \xib) (\Sb \cdot \etab) - \Lb
\cdot
        \Sb\right] = 4\omega (\Sb \cdot \etab) (\Sb \cdot \xib) \label{eq:3.20}
\\
  B^- & = & 4\omega \left[\hat N + 3 - (\Sb \cdot \xib) (\Sb \cdot \etab) + \Lb
\cdot
        \Sb\right] = 4\omega (\Sb' \cdot \xib) (\Sb' \cdot \etab).
\label{eq:3.21}
\end{eqnarray}
\par
%
%
It is now a question of finding the states $\phi^{\pm}_1$,~$\phi^{\pm}_2$ and
the
corresponding energies. This is facilitated by the fact that
$A^{\pm}$,~$B^{\pm}$
commute with the total number of quanta~$\hat N$ and angular momentum~$\Jb =
\Lb + \Sb$. Hence, we could express the states $\phi^{\pm}_1$,~$\phi^{\pm}_2$
in
terms of harmonic oscillator states with spin either~0 or~1, i.e., kets of the
type
\begin{equation}
  \left|N (l, s) jm\right\rangle = \sum_{\mu,\sigma} \left[ \left\langle
  l\mu, s\sigma | jm\right\rangle R_{Nl}(r) Y_{l\mu}(\theta,\varphi)
  \,\chi_{s,\sigma}\right]
\label{eq:3.22}
\end{equation}
where $s=1$, $\sigma=1$, 0,~$-1$, or $s=0$,~$\sigma=0$.\par
%
%
We note that $A^+$ and~$B^+$ commute with one another, as do also $A^-$
and~$B^-$, and thus rather than directly look for the eigenvalues of the
energy~$E$, we shall first consider the eigenstates of~$A^{\pm}$ and~$B^{\pm}$
and their corresponding eigenvalues, which we shall denote by~$\lambda^{\pm}_a$
and~$\lambda^{\pm}_b$, respectively.\par
%
%
We shall start our discussion by remarking that for fixed values of~$N$
and~$j$,
and for $s=0$, there is only one ket of the form~(\ref{eq:3.22}), whose parity
is~$(-1)^j$, namely
\begin{equation}
  \varphi_0(N) \equiv \left|N (j, 0) jm\right\rangle          \label{eq:3.23}
\end{equation}
where we introduced the shorthand notation~$\varphi_0(N)$ for this ket. Note
that
we indicate the explicit dependence of~$\varphi_0$ on~$N$, as this value will
change in the following section, but we do not include a dependence on~$j$ as
this
non-negative integer will remain fixed.\par
%
%
When the spin is~1, but the parity continues to be~$(-1)^j$, there is again
only one
ket, which is
\begin{equation}
  \varphi'_0(N) \equiv \left|N (j, 1) jm\right\rangle          \label{eq:3.24}
\end{equation}
where all states of spin~1 will be denoted by~$\varphi'$. On the other hand, if
$s=1$ but the parity is~$-(-1)^j$, there are then two kets
\begin{equation}
  \varphi'_{\pm}(N) \equiv \left|N (j\pm1, 1) jm\right\rangle.
\label{eq:3.25}
\end{equation}
\par
%
%
If $\lambda^{\pm}_a$, $\lambda^{\pm}_b$ can be determined in the various
instances discussed in the previous paragraph, then the matrix operators in
eqs.~(\ref{eq:3.15}), (\ref{eq:3.19}) become purely numerical matrices
\begin{equation}
  \left(\begin{array}{cc}
              \lambda^{\pm}_a - E^2 & 2E                                \\
              2E                                & \lambda^{\pm}_b - E^2 \\
           \end{array}\right)                    \label{eq:3.26}
\end{equation}
which will give rise to secular equations for the energy of the form
\begin{equation}
  \left(E^2 - \lambda^{\pm}_a\right) \left(E^2 - \lambda^{\pm}_b\right) - 4E^2
= 0.
  \label{eq:3.27}
\end{equation}
We shall now proceed to obtain these equations when the states are given
by~(\ref{eq:3.23}), (\ref{eq:3.24}), and~(\ref{eq:3.25}), respectively.\par
%
%
Let us start with the case of positive frequency~$\omega$ and spin~$s=0$.
{}From~(\ref{eq:3.16}), (\ref{eq:3.17}) and~(\ref{eq:3.23}), we obtain
\begin{equation}
  \lambda^+_a = 0 \qquad \lambda^+_b = 4\omega N              \label{eq:3.28}
\end{equation}
and thus the square of the energy in~(\ref{eq:3.27}) can take the values
\begin{equation}
  E^2 = 0 \qquad E^2 = 4 + 4\omega N.            \label{eq:3.29}
\end{equation}
The vanishing value gives rise to the phenomenon we have called a cockroach
nest~\cite{16}. So, in the following analysis, we shall disregard it and be
only
concerned with positive~$E^2$, such as the second expression
in~(\ref{eq:3.29}),
which gives an equally spaced spectrum for~$E^2$.\par
%
%
Turning now our attention to the state of spin~1 and parity~$(-1)^j$, we note
from~(\ref{eq:3.24}) that~\cite{17}
\begin{eqnarray}
  \left\langle N(l,1)jm | \Lb \cdot \Sb | N(l,1)jm \right\rangle & = &
\case{1}{2}
         \left[j(j+1) - l(l+1) - 2\right] \label{eq:3.30} \\
  \left\langle N(j,1)jm | (\etab \cdot \Sb) (\xib \cdot \Sb) | N(j,1)jm
\right\rangle
         & = & N + 1. \label{eq:3.31}
\end{eqnarray}
Thus from the middle expression in eqs.~(\ref{eq:3.16}), (\ref{eq:3.17}), we
obtain
that
\begin{equation}
  \lambda^+_a = 4\omega (N + 2) \qquad \lambda^+_b = 0      \label{eq:3.32}
\end{equation}
and the relevant square of the energy takes then the form
\begin{equation}
  E^2 = 4 + 4\omega (N + 2).             \label{eq:3.33}
\end{equation}
\par
%
%
For the case of spin~1 and parity~$-(-1)^j$, the kets are $\left|N (j\pm1,1) jm
\right\rangle$, and the non-vanishing matrix elements of~$\Lb \cdot \Sb$ are
given
by~(\ref{eq:3.30}), while those of $(\etab \cdot \Sb) (\xib \cdot \Sb)$ were
obtained in ref.~\cite{17}. Because of the existence of two kets instead of one
as
before, the operators~$A^+$ and~$B^+$ now become $2\times 2$ numerical
matrices. Their eigenvalues
\begin{equation}
  \lambda^+_a = 0 \qquad \lambda^+_b = 4\omega (N + 2)     \label{eq:3.34}
\end{equation}
correspond to the eigenstate
\begin{equation}
  \varphi'_1(N) = \left(\frac{(j+1) (N+j+2)}{(2j+1) (N+2)}\right)^{1/2}
  \varphi'_+(N) + \left(\frac{j (N-j+1)}{(2j+1) (N+2)}\right)^{1/2}
\varphi'_-(N)
  \label{eq:3.35}
\end{equation}
while the eigenvalues
\begin{equation}
  \lambda^+_a = 4\omega (N + 2) \qquad \lambda^+_b =  0    \label{eq:3.36}
\end{equation}
correspond to the eigenstate
\begin{equation}
  \varphi'_2(N) = \left(\frac{j (N-j+1)}{(2j+1) (N+2)}\right)^{1/2}
  \varphi'_+(N) - \left(\frac{(j+1) (N+j+2)}{(2j+1) (N+2)}\right)^{1/2}
\varphi'_-(N).
  \label{eq:3.37}
\end{equation}
In both cases, the relevant square of the energy is
\begin{equation}
  E^2 = 4 + 4\omega (N + 2)                 \label{eq:3.38}
\end{equation}
so from~(\ref{eq:3.33}) and~(\ref{eq:3.38}), we see that we have the same~$E^2$
for
all $s=1$~states.\par
%
%
In the case where the frequency is~$-\omega$, the analysis is entirely similar,
and only the results will be given. For~$s=0$ and parity~$(-1)^j$, we have
\begin{equation}
  \lambda^-_a = 0 \qquad \lambda^-_b = 4\omega (N + 3) \qquad E^2 = 4 + 4\omega
  (N + 3).                     \label{eq:3.39}
\end{equation}
For~$s=1$ and parity~$(-1)^j$, we get
\begin{equation}
  \lambda^-_a = 4\omega (N + 1) \qquad \lambda^-_b = 0  \qquad E^2 = 4 +
4\omega
  (N + 1).                     \label{eq:3.40}
\end{equation}
Finally, for~$s=1$ and parity~$-(-1)^j$, we have two possibilities
\begin{eqnarray}
  \lambda^-_a & = & 0 \qquad \lambda^-_b = 4\omega (N + 1) \qquad E^2 = 4 +
          4\omega (N + 1) \label{3.41} \\
  \lambda^-_a & = & 4\omega (N + 1) \qquad \lambda^-_b = 0 \qquad E^2 = 4 +
          4\omega (N + 1) \label{eq:3.42}
\end{eqnarray}
so again, for all situations when~$s=1$, we get the same~$E^2$.\par
%
%
The eigenkets for frequency~$-\omega$ will be denoted with a bar above, i.e.,
as
$\overline{\varphi}_0(N)$, $\overline{\varphi}'_0(N)$,
$\overline{\varphi}'_1(N)$,~$\overline{\varphi}'_2(N)$. An analysis similar to
the
one carried out between eqs.~(\ref{eq:3.23}) and~(\ref{eq:3.37}) shows that
\begin{equation}
  \overline{\varphi}_0(N) = \varphi_0(N) \qquad \overline{\varphi}'_0(N) =
  \varphi'_0(N)     \label{eq:3.43}
\end{equation}
while $\overline{\varphi}'_1(N)$,~$\overline{\varphi}'_2(N)$ are given by
\begin{eqnarray}
  \overline{\varphi}'_1(N) & = & \left(\frac{(j+1) (N-j+1)}{(2j+1)
(N+1)}\right)^{1/2}
           \varphi'_+(N) + \left(\frac{j (N+j+2)}{(2j+1) (N+1)}\right)^{1/2}
           \varphi'_-(N) \label{eq:3.44} \\
  \overline{\varphi}'_2(N) & = & \left(\frac{j (N+j+2)}{(2j+1)
(N+1)}\right)^{1/2}
           \varphi'_+(N) - \left(\frac{(j+1) (N-j+1)}{(2j+1)
(N+1)}\right)^{1/2}
           \varphi'_-(N) \label{eq:3.45}
\end{eqnarray}
with $\varphi_0(N)$, $\varphi'_0(N)$,~$\varphi'_{\pm}(N)$ given
by~(\ref{eq:3.23}),
(\ref{eq:3.24}), and~(\ref{eq:3.25}), respectively.\par
%
%
Note that whenever the eigenvalue of~$A^{\pm}$ is different from zero, that
of~$B^{\pm}$ vanishes, and viceversa. From the above analysis, it therefore
follows that $(4\omega)^{-1} \left(A^+ + B^+\right)$ has eigenvalue~$N$
if~$s=0$,
and~$N+2$ if~$s=1$, while for $(4\omega)^{-1} \left(A^- + B^-\right)$, it is
respectively $N+1$ for~$s=1$, and~$N+3$ for~$s=0$. All these eigenvalues are
displayed in table~\ref{tab:1}, where besides we give the spin and parity of
the
corresponding eigenkets.\par
%
%
The combined spectrum of the square of the energy, in units~$4\omega$, for both
frequencies~$\omega$ and~$-\omega$ is illustrated in figure~\ref{fig:2}, where
below the level columns we indicate the value of the spin and the sign of the
frequency. The spectrum shows an extraordinary degeneracy, which we intend to
explain in the next section through consideration of supersymmetry that
eventually
leads to an appropriate superalgebra.\par
%
%
\section{Supersymmetry and superalgebra for the two-body system with a new
type of Dirac oscillator interaction}
\setcounter{equation}{0}
As we showed in the previous section that all states of spin~1, regardless of
whether they have parity~$(-1)^j$ or~$-(-1)^j$, give rise to the same
eigenvalue~$N+2$ for~$A^+ + B^+$, and~$N+1$ for~$A^- + B^-$, it is convenient
to
group them as vectors in the notation
\begin{eqnarray}
  \varphib'(N) & = & \left(\begin{array}{c}
                                           \varphi'_0(N) \\[0.1cm]
                                           \varphi'_1(N) \\[0.1cm]
                                           \varphi'_2(N)
                                        \end{array}\right) \nonumber \\[0.1cm]
  \overline{\varphib}'(N) & = & \left(\begin{array}{c}

\overline{\varphi}'_0(N) \\[0.1cm]

\overline{\varphi}'_1(N) \\[0.1cm]

\overline{\varphi}'_2(N)
                                                         \end{array}\right)
  \label{eq:4.1}
\end{eqnarray}
where the components in the unbarred and barred vectors were all given
explicitly
in section~2.\par
%
%
We remarked previously that $(4\omega)^{-1} \left(A^+ + B^+\right)$, when
applied
to $\varphi_0(N)$,~$\varphib'(N)$, gives the eigenvalues $N$,~$N+2$,
respectively,
while $(4\omega)^{-1} \left(A^- + B^-\right)$, applied to
$\overline{\varphi}_0(N)$,~$\overline{\varphib}'(N)$, gives $N+3$,~$N+1$. In
analogy to the superstate discussed in ref.~\cite{4} for the one-particle
problem,
let us now introduce a superstate for the two-body problem, defined by the
expression
\begin{equation}
  \Psi = \left(\begin{array}{c}
                         \varphi_0(N)                      \\[0.1cm]
                         \overline{\varphib}'(N-1)   \\[0.1cm]
                         \varphib'(N-2)                    \\[0.1cm]
                         \overline{\varphi}_0(N-3)
                      \end{array}\right)         \label{eq:4.2}
\end{equation}
with $\varphi_0$, $\overline{\varphi}_0$, $\varphib'$,~$\overline{\varphib}'$
defined by~(\ref{eq:3.23}), (\ref{eq:3.43}), and~(\ref{eq:4.1}), respectively.
{}From
the discussion of the previous section, and the definitions (\ref{eq:3.16}),
(\ref{eq:3.17}), (\ref{eq:3.20}), and~(\ref{eq:3.21}) of $A^{\pm}$,~$B^{\pm}$,
it is
clear that~$\Psi$ is an eigenstate with eigenvalue~$N$ of the following
supersymmetric Hamiltonian
\begin{equation}
  H \Psi \equiv \left(\begin{array}{cccc}
                                    H^+ & 0    & 0    & 0   \\[0.1cm]
                                    0    & H^- & 0    & 0   \\[0.1cm]
                                    0    & 0    & H^+ & 0   \\[0.1cm]
                                    0    & 0    & 0    & H^-
                                \end{array}\right)
                        \left(\begin{array}{c}
                                    \varphi_0(N)                      \\[0.1cm]
                                    \overline{\varphib}'(N-1)   \\[0.1cm]
                                    \varphib'(N-2)                    \\[0.1cm]
                                    \overline{\varphi}_0(N-3)
                                 \end{array}\right)
  = N \left(\begin{array}{c}
                    \varphi_0(N)                      \\[0.1cm]
                    \overline{\varphib}'(N-1)   \\[0.1cm]
                    \varphib'(N-2)                    \\[0.1cm]
                    \overline{\varphi}_0(N-3)
                \end{array}\right)          \label{eq:4.3}
\end{equation}
where
\begin{eqnarray}
  H^+ & = & (4\omega)^{-1} \left(A^+ + B^+\right) = (\Sb \cdot \xib) (\Sb \cdot
\etab)
             + (\Sb' \cdot \etab) (\Sb' \cdot \xib)        \label{eq:4.4} \\
  H^- & = & (4\omega)^{-1} \left(A^- + B^-\right) = (\Sb \cdot \etab) (\Sb
\cdot \xib)
             + (\Sb' \cdot \xib) (\Sb' \cdot \etab).        \label{eq:4.5}
\end{eqnarray}
\par
%
%
The problem now resides in finding the corresponding supercharges, i.e., the
counterparts of the operators $Q$,~$Q^{\dagger}$ of the one-body case given in
ref.~\cite{4}. We shall retain the notation
$Q$,~$Q^{\dagger}$, with the latter being the Hermitian conjugate of the
former.
We would like of course to recreate the relation~(3.2) in ref.~\cite{4}, i.e.,
\begin{equation}
  \left\{Q, Q^{\dagger}\right\} \equiv Q Q^{\dagger} + Q^{\dagger} Q = H
  \label{eq:4.6}
\end{equation}
but also that $Q$, when applied to~$\Psi$, would relate $\varphi_0(N)$
with~$\overline{\varphib}'(N-1)$, $\overline{\varphib}'(N-1)$
with~$\varphib'(N-2)$, and the latter with~$\overline{\varphi}_0(N-3)$.\par
%
%
It is obvious that $Q$ must contain the annihilation operator~$\xib$, but in a
form
that should commute with the total angular momentum $\Jb = \Lb + \Sb$, as
$j$,~$m$ are the same for all states in~$\Psi$. This last point can only be
achieved if we consider the scalar products of~$\xib$ with the spin operators
of
the two particles or, better still, with their sum and difference
$\Sb$,~$\Sb'$,
defined in~(\ref{eq:3.18}). Thus $Q$ must be a $4\times4$ matrix of the type
appearing in~(\ref{eq:4.3}) for~$H$, but whose elements depend on $(\Sb \cdot
\xib)$,~$(\Sb' \cdot \xib)$. We note though from standard Racah
algebra~\cite{18}
that $\Sb'$ can connect only states of spin~0 with those of spin~1 or
viceversa,
while $\Sb$ can only connect states with the same spin. As
$\varphi_0$,~$\overline{\varphi}_0$ have spin~0 while $\varphib'$,
{}~$\overline{\varphib}'$ have spin~1, this immediately suggests that
\begin{equation}
  Q = \Deltab^{\dagger} \cdot \xib \qquad
  \Deltab^{\dagger} \equiv \left(\begin{array}{cccc}
                                                     0     & 0     & 0    & \Sb
\\
                                                     \Sb' & 0    & 0     & 0
\\
                                                     0     & \Sb & 0     & 0
\\
                                                     0     & 0    & \Sb' & 0
                                                  \end{array}\right).
  \label{eq:4.7}
\end{equation}
In turn, the Hermitian conjugate~$Q^{\dagger}$ is given by
\begin{equation}
  Q^{\dagger} = \Deltab \cdot \etab \qquad
  \Deltab \equiv
        \left(\begin{array}{cccc}
                    0     & \Sb' & 0    & 0       \\
                    0     & 0     & \Sb & 0       \\
                    0     & 0     & 0     & \Sb'  \\
                    \Sb  & 0    & 0     & 0
                 \end{array}\right).
  \label{eq:4.8}
\end{equation}
\par
%
%
{}From the relation
\begin{equation}
  (\Sb \cdot \xib) (\Sb' \cdot \xib) = (\Sb \cdot \etab) (\Sb' \cdot \etab) = 0
  \label{eq:4.9}
\end{equation}
we immediately get
\begin{equation}
  \left\{Q, Q\right\} = \left\{Q^{\dagger}, Q^{\dagger}\right\} = 0 \qquad
  \left\{Q, Q^{\dagger}\right\} = H         \label{eq:4.10}
\end{equation}
from which it follows that
\begin{equation}
  \left[H, Q\right] = \left[H, Q^{\dagger}\right] = 0.         \label{4.11}
\end{equation}
We have therefore found for the two-body problem the su(1/1)
superalgebra~\cite{14a} characteristic of supersymmetric quantum
mechanics~\cite{4a}, and analogous to the one-body relations obtained in
eq.~(3.2)
of ref.~\cite{4}.\par
%
%
As a last point, we would like to obtain the superalgebra responsible for the
degeneracy we observe in figure~1 for our two-body problem with a Dirac
oscillator interaction. For this purpose, we note that
$\Deltab$,~$\Deltab^{\dagger}$ are vector matrices of components
\begin{equation}
  \Delta_i, \Delta^{\dagger}_j \qquad i, j = 1, 2, 3       \label{eq:4.12}
\end{equation}
and in the appendix we prove that the latter satisfy the following
anticommutation
relations
\begin{equation}
  \left\{\Delta_i, \Delta_j\right\} = \left\{\Delta^{\dagger}_i,
              \Delta^{\dagger}_j\right\} = 0 \qquad
  \left\{\Delta_i, \Delta^{\dagger}_j\right\} = \delta_{ij}\, \I
\label{eq:4.13}
\end{equation}
where $\I$ is a $4\times4$ unit matrix. Equation~(\ref{eq:4.13}) means that
$\Delta^{\dagger}_i$,~$\Delta_j$ behave as fermion creation and annihilation
operators.\par
%
%
On the other hand, $\eta_i$,~$\xi_j$ are boson creation and annihilation
operators,
which satisfy the following commutation relations
\begin{equation}
  \left[\eta_i, \eta_j\right] = \left[\xi_i, \xi_j\right] = 0 \qquad
  \left[\xi_i, \eta_j\right] = \delta_{ij}           \label{4.14}
\end{equation}
and of course, from their very definition, the boson $\eta_i$,~$\xi_j$ and
fermion
$\Delta^{\dagger}_i$,~$\Delta_j$ operators commute among themselves.\par
%
%
The supersymmetric Hamiltonian, in the form~(\ref{eq:4.10}), can now be written
as
\begin{equation}
  H = \left(\eta_i \Delta_i\right) \left(\xi_j \Delta^{\dagger}_j\right) +
  \left(\xi_j \Delta^{\dagger}_j\right) \left(\eta_i \Delta_i\right)
  \label{eq:4.15}
\end{equation}
where repeated indices are summed over $i$,~$j=1$, 2,~3. By using all the
above-mentioned commutation and anticommutation relations, and the property
\begin{equation}
  \Sb'^2 = 3 - \Sb^2               \label{eq:4.16}
\end{equation}
it is straightforward to show that
\begin{equation}
  H = \etab \cdot \xib \,\I + \Deltab^{\dagger} \cdot \Deltab = \left(
  \begin{array}{cccc}
     \etab\cdot\xib+\Sb^2 & 0                                     & 0
           & 0                                     \\
     0                                 & \etab\cdot\xib+3-\Sb^2 & 0
           & 0                                     \\
     0                                 & 0
& \etab\cdot\xib+\Sb^2
           & 0                                     \\
     0                                 & 0
& 0
           & \etab\cdot\xib+3-\Sb^2
  \end{array}\right).          \label{eq:4.17}
\end{equation}
\par
%
%
It is then obvious that all the operators
\begin{eqnarray}
  \C_{ij} & = & \eta_i \xi_j \,\I \qquad C_{ij} = \Delta^{\dagger}_i \Delta_j
          \label{eq:4.18} \\
  T_{ij} & = & \eta_i \Delta_j \qquad U_{ij} = \Delta^{\dagger}_i \xi_j
          \label{eq:4.19}
\end{eqnarray}
commute with~$H$. The operators~(\ref{eq:4.18}) and~(\ref{eq:4.19}) are the
even
and odd generators of a symmetry superalgebra respectively, as they satisfy the
following commutation and anticommutation relations
\begin{eqnarray}
  \left[\C_{ij}, \C_{i'j'}\right] & = & \delta_{ji'} \C_{ij'} - \delta_{ij'}
\C_{i'j}
           \label{eq:4.20} \\
  \left[C_{ij}, C_{i'j'}\right] & = & \delta_{ji'} C_{ij'} - \delta_{ij'}
C_{i'j}
           \label{eq:4.21} \\
  \left[\C_{ij}, T_{i'j'}\right] & = & \delta_{ji'} T_{ij'}  \qquad
           \left[\C_{ij}, U_{i'j'}\right] = - \delta_{ij'} U_{i'j}
\label{eq:4.22} \\
  \left[C_{ij}, T_{i'j'}\right] & = & - \delta_{ij'} T_{i'j}  \qquad
           \left[C_{ij}, U_{i'j'}\right] = \delta_{ji'} U_{ij'}
\label{eq:4.23} \\
  \left\{T_{ij}, T_{i'j'}\right\} & = & 0  \qquad \left\{U_{ij},
U_{i'j'}\right\} = 0
           \label{eq:4.24} \\
  \left\{T_{ij}, U_{i'j'}\right\} & = & \delta_{ji'} \C_{ij'}  + \delta_{ij'}
C_{i'j}.
           \label{eq:4.25}
\end{eqnarray}
The latter are the defining relations of a u(3/3) superalgebra~\cite{14a}.\par
%
%
Such a superalgebra fully explains the accidental degeneracy of the levels of
the
two-body system. It indeed contains some operators giving rise to transitions
between degenerate eigenstates of a given component of the supersymmetric
Hamiltonian~(\ref{eq:4.17}) (namely, the even generators~$\C_{ij}$
and~$C_{ij}$),
as well as some operators generating transitions between degenerate eigenstates
of different components of the same (namely, the odd generators~$T_{ij}$
and~$U_{ij}$).\par
%
%
One should remark that the su(1/1) superalgebra of supersymmetric quantum
mechanics is embedded into~u(3/3), since the operators~$H$, $Q$,
and~$Q^{\dagger}$, defined in~(\ref{eq:4.17}), (\ref{eq:4.7}),
and~(\ref{eq:4.8}), can
be rewritten as
\begin{equation}
  H = \sum_i \left(\C_{ii} + C_{ii}\right) \qquad Q = \sum_i U_{ii} \qquad
  Q^{\dagger} = \sum_i T_{ii}.         \label{4.26}
\end{equation}
\par
%
%
\section{Conclusion}
The objective mentioned at the beginning of the paper of finding the symmetry
superalgebra of the two-body system with a new type of Dirac oscillator
interaction has been achieved. Moreover, we did identify such a superalgebra
with
u(3/3), and we did show that it contains the su(1/1) superalgebra of
supersymmetric quantum mechanics as a subsuperalgebra.\par
%
%
This u(3/3) superalgebra plays the same role for the present problem as the
algebra~u(3) for the standard harmonic oscillator. For the latter, it is well
known
that by adding the operators~$\eta_i \eta_j$, $\xi_i \xi_j$, $i$,~$j=1$, 2,~3,
to
the $u(3)$~generators $\eta_i \xi_j$, $i$,~$j=1$, 2,~3, one obtains a dynamical
Lie algebra of the type sp(6,\R)~\cite{19}.\par
%
%
Can one find in a similar way a dynamical Lie superalgebra for the two-body
system with a new type of Dirac oscillator interaction? The answer is of course
yes, as one only has to add to the u(3/3) generators~$\C_{ij}$, $C_{ij}$,
$T_{ij}$,
$U_{ij}$, $i$,~$j=1$, 2,~3, of eqs.~(\ref{eq:4.18}), (\ref{eq:4.19}), the
operators
mentioned in the previous paragraph, plus operators of the form~$\eta_i
\Delta^{\dagger}_j$, $\xi_i \Delta_j$, $\Delta^{\dagger}_i \Delta^{\dagger}_j$,
$\Delta_i \Delta_j$, $i$,~$j=1$, 2,~3. It is indeed straightforward to show
that
the whole set of operators generate a dynamical superalgebra of the type
osp(6/6,\R)~\cite{14a}.\par
%
%
\section*{Acknowledgments}
One of the authors (CQ) would like to thank the Instituto de F\'\i sica of the
UNAM
for its kind hospitality during a stay. A grant from the Minist\`ere de
l'Education
Nationale et de la Culture, Communaut\'e Fran\c caise de Belgique, is also
acknowledged.\par
\newpage
%
%
\section*{Appendix. Some relations for spin operators}
\renewcommand{\theequation}{A.\arabic{equation}}
\setcounter{section}{0}
\setcounter{equation}{0}
By using the well-known relation
\begin{equation}
  \sigma_i \sigma_j = \delta_{ij} + i \epsilon_{ijk} \sigma_k
\label{eq:2.7}
\end{equation}
for the Pauli spin matrices, it is easy to check that the spin operators~$\Sb$
and~$\Sb'$, defined in~(\ref{eq:3.18}), satisfy the following relations
\begin{eqnarray}
  S_i S'_j + S_j S'_i & = & 0              \label{eq:A.1}  \\
  S'_i S_j + S'_j S_i & = & 0              \label{eq:A.2}  \\
  S'_i S'_j + S_j S_i & = & \delta_{ij}.             \label{eq:A.3}
\end{eqnarray}
Then eqs.~(\ref{eq:4.9}) and~(\ref{eq:4.16}) directly follow
from~(\ref{eq:A.1})
and~(\ref{eq:A.3}), respectively.\par
%
%
The proof of eq.~(\ref{eq:4.13}) is also straightforward. Considering first the
second anticommutator in~(\ref{eq:4.13}), we obtain
\begin{eqnarray}
  \lefteqn{\left\{\Delta^{\dagger}_i, \Delta^{\dagger}_j\right\}} \nonumber \\
  & = & \left(\begin{array}{cccc}
                       0     & 0    & 0     & S_i \\
                       S'_i & 0    & 0     & 0    \\
                       0     & S_i & 0     & 0    \\
                       0     & 0    & S'_i & 0
                       \end{array}\right)
           \left(\begin{array}{cccc}
                       0     & 0     & 0     & S_j \\
                       S'_j & 0     & 0     & 0    \\
                       0     & S_j & 0      & 0    \\
                       0     & 0     & S'_j & 0
                   \end{array}\right)
        + \left(\begin{array}{cccc}
                       0     & 0     & 0     & S_j \\
                       S'_j & 0    & 0      & 0    \\
                       0     & S_j & 0      & 0    \\
                       0     & 0     & S'_j & 0
                   \end{array}\right)
          \left(\begin{array}{cccc}
                       0     & 0    & 0     & S_i \\
                       S'_i & 0    & 0     & 0    \\
                       0     & S_i & 0     & 0    \\
                       0     & 0    & S'_i & 0
                   \end{array}\right) \nonumber \\
  & = & \left(\begin{array}{cccc}
                       0                          & 0
& S_iS'_j + S_jS'_i
                               & 0                          \\
                       0                          & 0
& 0
                               & S'_iS_j + S'_jS_i \\
                       S_iS'_j + S_jS'_i & 0                          & 0
                               & 0                          \\
                       0                          & S'_iS_j + S'_jS_i & 0
                               & 0
                    \end{array}\right) = 0                \label{eq:A.4}
\end{eqnarray}
by successively using~(\ref{eq:4.7}), (\ref{eq:A.1}), and~(\ref{eq:A.2}).
Similarly, we
get $\left\{\Delta_i, \Delta_j\right\} = 0$ for the first anticommutator
in~(\ref{eq:4.13}). Finally, the last anticommutator follows
from~(\ref{eq:4.7}),
(\ref{eq:4.8}), and~(\ref{eq:A.3}):
\begin{eqnarray}
  \lefteqn{\left\{\Delta^{\dagger}_i, \Delta_j\right\}} \nonumber\\
  & = & \left(\begin{array}{cccc}
                       0     & 0    & 0     & S_i \\
                       S'_i & 0    & 0     & 0    \\
                       0     & S_i & 0     & 0    \\
                       0     & 0    & S'_i & 0
                    \end{array}\right)
           \left(\begin{array}{cccc}
                       0     & S'_j & 0     & 0     \\
                       0     & 0     & S_j  & 0     \\
                       0     & 0     & 0     & S'_j \\
                       S_j & 0     & 0      & 0
                    \end{array}\right)
         + \left(\begin{array}{cccc}
                       0     & S'_j & 0     & 0    \\
                       0     & 0     & S_j & 0     \\
                       0     & 0     & 0     & S'_j \\
                       S_j & 0     & 0     & 0
                    \end{array}\right)
           \left(\begin{array}{cccc}
                       0     & 0    & 0     & S_i \\
                       S'_i & 0    & 0     & 0    \\
                       0     & S_i & 0     & 0    \\
                       0     & 0    & S'_i & 0
                    \end{array}\right) \nonumber \\
  & = & \left(\begin{array}{cccc}
                       S_iS_j + S'_jS'_i & 0                           & 0
                               & 0                          \\
                       0                          & S'_iS'_j + S_j S_i & 0
                               & 0                          \\
                       0                          & 0
& S_iS_j + S'_jS'_i
                               & 0                          \\
                       0                          & 0
& 0
                               & S'_iS'_j + S_jS_i
                    \end{array}\right) = \delta_{ij}\,\I.
\label{eq:A.5}
\end{eqnarray}
\par
\newpage
%
%

\newpage
%
%
\section*{Tables and table captions}
\begin{table}[h]

\caption{The eigenvalues of the operators~$A^+$, $B^+$, $A^-$,~$B^-$, and of
their
sums~$A^+ + B^+$, $A^- + B^-$, are indicated in units of~$4\omega$. The
spin~$s=0$
or~1 and parity~${\cal P} = (-1)^j$ or $-(-1)^j$ of the states are also shown
in
the first two colums.}\label{tab:1}

\vspace{1cm}
\begin{tabular}{llllllll}
  \hline\\[-0.2cm]
  $s$ & $\m\mathcal{P}$ & $A^+$ & $B^+$  & $A^-$  & $B^-$  & $A^++B^+$ &
$A^-+B^-$
        \rule[-0.3cm]{0cm}{0.6cm}\\[0.2cm]
  \hline\\[-0.2cm]
  0 & $\m(-1)^j$ & 0        & $N$     & 0         & $N+3$  & $N$     & $N+3$
\\[0.2cm]
  1 & $\m(-1)^j$ & $N+2$ & 0        & $N+1$  & 0         & $N+2$ & $N+1$
\\[0.2cm]
  1 & $-(-1)^j$    & 0        & $N+2$ & 0         & $N+1$  & $N+2$ & $N+1$
\\[0.2cm]
  1 & $-(-1)^j$    & $N+2$ & 0        & $N+1$  & 0         & $N+2$ & $N+1$
\\[0.2cm]
  \hline
\end{tabular}
\end{table}
\newpage
%
%
\section*{Figure captions}
\begin{figure}[h]
\caption{Square of the energy levels in units~$4\omega$ for the two-body Dirac
oscillator for frequencies~$+\omega$ and~$-\omega$, with the sign indicated in
the abscissa as also the spin values~$s=0$ or~1. The action of the supercharge
operators $Q$,~$Q^{\dagger}$ is also shown.}\label{fig:2}
\end{figure}

\end{document}